\documentclass{article}

\usepackage{graphicx}
\usepackage{amsmath}
\usepackage{amsfonts}
\usepackage{amssymb}
\usepackage{natbib}

\begin{document}


\title{Fluctuations and \ Dissipation of Coherent Magnetization}
\author{A. Rebei and G.J. Parker \\
Seagate Research Center, Pittsburgh, PA 15222 }
\date{July, 2002 }

\maketitle

\begin{abstract}
A quantum mechanical model is used to derive a generalized 
   Landau-Lifshitz equation for a magnetic moment, including 
fluctuations and 
dissipation. \ The model reproduces the Gilbert-Brown form of the
equation in the classical limit. \ The magnetic moment is linearly 
coupled to a 
reservoir of bosonic degrees of freedom. \ Use of generalized coherent states makes the semiclassical
limit more transparent within a path-integral formulation. \ A general
fluctuation-dissipation theorem is derived. \ The magnitude of the magnetic
moment also fluctuates beyond the Gaussian approximation. \ We 
 discuss how the approximate stochastic description of the thermal field 
follows from our 
result. \ As an example, we go beyond the linear-response method 
and show how the thermal fluctuations become
 anisotropy-dependent even in the uniaxial case. 
\end{abstract}

\bigskip
\bigskip

{\text{ PACS: 75.10.Jm, 75.30.Gw, 76.60.Es }}
\bigskip

\newpage


\newpage

\section{INTRODUCTION}

\bigskip The study of thermally induced magnetization reversal was 
first carried out by 
Brown \cite{brown}. \ His approach was to introduce a noise term into the 
Landau-Lifshitz 
equation of motion, essentially constructing a Langevin-type 
equation, which we call here 
the Landau-Lifshitz-Gilbert-Brown equation (LLGB). \ From the 
LLGB equation the Fokker-Plank equation can be derived that 
describes the time
 evolution of the  probability density distribution of the moment 
orientations. \ Solution 
of this problem was carried out by Brown for the case of an axially symmetric
 potential and later by Coffey et al \cite{coffey} for non-axially symmetric 
cases. \ The numerical solution of the Langevin equation was used by Lyberatos 
et al \cite{lyb}, and has since been applied to the study of magnetization
 reversal by a number of authors, \cite{nakatani,safonov}. \ Recently, 
Wang et al \cite{wang} have developed an approach introducing a tensor 
form of the damping constant and applied this to the calculation of first 
mean passage time in the case of an elongated grain represented as a chain 
of coupled particles. \ Many of these calculations are motivated by the need 
to understand high frequency magnetization reversal processes in magnetic 
recording. \ The process of reading information currently involves 
GMR sensors, 
the size of which is continually reducing as recording densities 
increase. \ This 
led Smith and Arnett \cite{NeilSmith} to the conjecture that noise due to 
magnetization fluctuations in the read head would be a limiting factor on the 
device size. \ This is clearly an important problem, which has been further 
developed by Smith \cite{smith} and Bertram and co-workers \cite{bertram}, who 
have also studied the full micromagnetic description of the problem using an 
approach in which the thermal noise is distributed among the normal 
modes \cite{bertram2}. 

\bigskip Clearly the introduction of thermal fluctuations in the micromagnetic 
formalism is  important both from the point of view of the physics of 
magnetization processes and also in relation to important practical 
problems of magnetic recording. \ Central to all models, both analytical 
and numerical, is the introduction of a magnetization fluctuation or 
an effective field via the fluctuation-dissipation theorem (FDT) 
\cite{callen_and_welton}. \ The FDT has a strong physical justification, as 
discussed in detail by Landau and Lifshitz \cite{ll}, however it should be 
stressed that it is strictly valid only for small fluctuations about the local 
minimum. \ A more serious problem with the use of the Langevin equation in 
the LLGB 
form is the dissipative term itself, which has no microscopic 
justification. \ It 
is clearly important to understand the whole problem of 
fluctuations and dissipation 
within a first priciples quantum mechanical approach, if the limits of 
the current 
models are to be established and more fundamental theoretical approaches 
derived.  The demand for higher density recording media 
and faster switching rates requires the use  
of structures on the
nanometer scale or less. \ Quantum mechanical 
effects are then bound to become more
and more important to consider in these systems. \ Effects such 
as magneto-optical interactions may even invalidate the simple
damping term that is currently used in the Landau-Lifshitz equation. \ This
prompted us to investigate whether the LLGB equation 
can be recovered from a more
fundamental treatment rather than the ad hoc approach presently used. \ Hence
it seems natural to ask in what limit the LLGB equation can be recovered
starting from a quantum model.

\bigskip
 \ In this paper, we make a small step toward that goal. \
 To address the above questions, we take a simple quantum 
model, that of a single
particle with large spin interacting with a heat bath and an external 
magnetic field.
\ The spin is taken to be large since we are primarily 
interested in a semiclassical
representation of the magnetization vector. \ This simple model is sufficient
 to
allow us to study the effects of thermal fluctuations in many different 
cases, such as  magnetization switching in a single
domain magnetic particle with uniaxial anisotropy, an important problem in
magnetic recording physics. \ The bath is taken to be of bosonic nature. \
Nothing else needs to be 
 assumed to enable us to include various mechanisms
of interaction between the magnetic moment and the 
environment. \ We calculate the equations 
of motion of the magnetization and that of an associated
 fluctuating field in the semi-classical limit. \ Since 
our interest is mainly in the semi-classical limit, coherent 
states (CS) are
the natural choice for the representation of the system. \ These 
states have the
property of minimizing the Heisenberg uncertainty relations. \ Furthermore,
the calculations are based on expressing the density matrix of the 
particle-bath system in terms of
path-integrals as in the Feynman-Vernon formalism 
\cite{Feynman,Feynman2,Caldeira}. \ Clearly this method allows 
a consistent treatment of the magnetization 
and fluctuations from the start.\ If the thermal field
is decoupled from the magnetization, 
 the LLGB
equation will be shown to 
correspond to a given choice of density of states of the
reservoir and of its interaction parameters with the magnetic moment. \ 

\bigskip \ This work is able to provide a different
angle from which to discuss the LLGB equation and its limitation. \ We 
also set a basis against which we can examine the  discrepancy in the 
recent calculations of the noise spectrum in magnetic recording heads   
\cite{NeilSmith,bertram2}. \  Therefore,  a treatment 
of the noise problem by a self-consistent method, such as 
the one presented here, can shed some light on why 
this difference exists.  \ We stress that our results are 
very general for the model considered and no linear approximation 
is assumed.

\bigskip  The paper is organized as follows. \ In section two, \ we 
introduce a simple model Hamiltonian that can describe
dissipation and fluctuations. \ We linearly couple a  single domain
magnetic particle to an external magnetic field and to a Bosonic bath with
infinite number of 
degrees of freedom. \ In section three, we show how to calculate the
reduced density matrix elements of the magnetic moment. \ The density matrix
elements are naturally expressed in terms of path-integrals over the
configuration space of the moment \cite{Feynman2}. \ In section four, we 
derive coupled equations for the magnetization and fluctuations.  \ We 
show that 
 a general
fluctuation-dissipation theorem is satisfied. \ We also 
demonstrate how to recover the LLGB equation by 
decoupling the thermal fluctuations, taking the high temperature
limit and constraining the choice of reservoir. \ In the particular case of
LLGB, this corresponds to Gaussian fluctuations and constant dissipation.
\ Similar results have been obtained for the case when the magnetic moment is
replaced by a harmonic oscillator. \ This is no surprise since in this case,
the semi-classical approximation corresponds to a particle with large spin.
\ In section five, we compare the classical stochastic treatment to this
quantum treatment. \ As an example we calculate the fluctuating field for a
single domain particle with the external field directed along the easy axis as
a function of the anisotropy constant. \ Finally in the last section, we
summarize our results and state 
  results based on our derived equations 
when generalized to the anisotropic case.

\bigskip

\section{ DEFINITION OF THE MODEL   }

\bigskip

\indent The model we choose is simple but general enough to include many 
interesting
physical situations. \ It is mainly motivated by the recent work
of Safonov and Bertram \cite{Safonov}. \ They used a two-level impurity system
to simulate relaxation effects in a single domain grain. \ They showed that
the damping in their model is of the Gilbert form. No fluctuations are
considered in their calculation. If we consider a collection of spins that are
independent, then the magnetization vector, $\mathbf{M}$, is a simple sum of
these coherent spins,
\begin{equation}
\mathbf{M}=g\mu_{B}\frac{ N \mathbf{S}}{V},
\end{equation}
where $\mathbf{S}$ is the spin vector and ${g\mu_{B}}/{\hbar}$ 
is the gyromagnetic ratio. \
$\mu_{B} $ is the Bohr magneton and $V$ is the volume of the system. In the
following we set $\hbar=1$, 
$g\mu_{B}=1$ and the density $\frac{N}{V}=1$. \ From now 
on, we use the words spin and
magnetic moment interchangeably.

\bigskip

We take a single spin $\mathbf{S}$ $\left( \mathbf{S}^{2}\gg1\right)  $ and
couple it linearly to a set of oscillators and to a constant external field
$\mathbf{H}$. \ The former may represent phonons, a time-dependent magnetic
field or other Bosonic degrees of freedom. \ No assumption will be made about
the coupling constants or the density of states of the reservoir. \ The
Hamiltonian assumes the following form:
\begin{equation}
\mathcal{H}=-\mathbf{H\cdot S}+\sum_{k}\omega_{k}a_{k}^{+}a_{k}+\sum_{k}%
\gamma_{k}a_{k}^{+}S_{-}+\sum_{k}\gamma_{k}^{\ast}S_{+}a_{k}
\label{hamiltonian}%
\end{equation}
where $\mathbf{H}$ is a static external magnetic field. \ $\mathbf{S}$ is the
magnetic moment of a single particle. \ $a_{k}^{+}$\ and $a_{k}$\ are creation
and annihilation operators of the reservoir. \ The coupling constants 
$\gamma_{k}$  may be time-dependent, but will be taken 
as independent of time in the final result. \ The field $\mathbf{H}$\ is taken
along the z-axis, the axis of spin quantization. \ Coupling the z-component of
the vector $\mathbf{S}$ to the reservoir can be easily added, but it will be
omitted in this work. \ This Hamiltonian is sufficient to describe all the
desired
physics. \ Using the equation of motion for $S_{z}$, it is
trivial to see that it is not a constant of the motion and hence no
linearization is implied in this model.

\bigskip

The operators are in the Heisenberg representation. \ The spin operator S
satisfies the usual commutation relations \cite{Edmonds}
\begin{equation}
\left[  \mathbf{S}^{2},S_{\pm}\right]  =0
\end{equation}
with
\begin{equation}
\mathbf{S}^{2}=\frac{1}{2}\left\{  S_{+},S_{-}\right\}  +S_{z}^{2},
\end{equation}
where the curly brackets are for anticommutation and
\begin{equation}
S_{+}=S_{x}+iS_{y}%
\end{equation}%
\begin{equation}
S_{-}=S_{x}-iS_{y}%
\end{equation}
while the operators of the reservoir satisfy Bose commutation relations,%

\begin{equation}
\left[  \ a_{k},a_{k^{\prime}}^{+}\right]  =\delta_{kk^{\prime}} \; .%
\end{equation}
Instead of the usual Fock space representation, we use a CS space
representation for these operators \cite{Glauber,Blaizot,Negele-Orland} .

\bigskip

\indent For a harmonic oscillator with position $q_{k}$, momentum $p_{k}$ and
frequency $\omega_{k}$, the CS $\left|  \Phi_{k}\right\rangle $ are
defined as eigenfunctions of the annihilation operator $a_{k}=\left(
\frac{\omega_{k}}{2}\right)  ^{1/2}\left(  q_{k}+\frac{i}{\left(  2\omega
_{k}\right)  ^{1/2}}p_{k}\right)  $%

\begin{equation}
a_{k}|\mathbf{\Phi}_{k}\rangle=\Phi_{k}|\mathbf{\Phi}_{k}\rangle
\end{equation}
with complex eigenvalues, $\Phi_{k}$ \cite{Glauber}. \ These states can also be generated
from the ground state by applying a displacement operator $D\left(
z_{k}\right)  $ which defines a one-to-one correspondence between the complex
plane and the oscillator states,
\begin{equation}
\left|  z_{k}\right\rangle =D\left(  z_{k}\right)  \left|  0_{k}\right\rangle
\end{equation}
and
\begin{equation}
D\left(  z_{k}\right)  =\exp\left(  z_{k}a_{k}^{+}-z_{k}^{\ast}a_{k}\right)  .
\end{equation}
CS's form an overcomplete basis and satisfy the minimum uncertainty
relation. Hence they are the most suitable representation for a semi-classical
treatment. \ We also adopt the normalization in \cite{Blaizot}
\begin{equation}
\langle\Phi_{k}|\Phi_{k}^{\prime}\rangle=e^{\Phi_{k}^{\ast}\Phi_{k}^{\prime}} .%
\end{equation}
They also satisfy the following relation, the resolution of the identity
operator,
\begin{equation}
\int\frac{d\Phi_{k}^{\ast}d\Phi_{k}}{2\pi i}e^{-\Phi_{k}^{\ast}\Phi_{k}}%
|\Phi_{k}\rangle\langle\Phi_{k}|=1.
\end{equation}
The latter relation is essential for a path-integral representation in terms
of CS's.

\bigskip

Similarly for the spin states, we use a CS
representation \cite{Radcliffe,Permov}. \ They are defined by analogy to
the harmonic oscillator CS's. \ The spin components in this state
satisfy a minimum uncertainty relation, i.e., two of the three components
commute \cite{Arrechi}. As in the harmonic oscillator case, a `ground' state
$ |0 \rangle$ is required from which to generate all the other states. \ In this case the
state with the largest $S_{z}$ component is taken as the reference state. \ If the
z-axis is taken as the quantization axis and if we take $\mathbf{S}%
^{2}=j\left(  j+1\right)  $, then by definition, we have%

\begin{equation}
|0\rangle \equiv |j,j\rangle,
\end{equation}
and
\begin{equation}
S_{z}|0\rangle=j|0\rangle
\end{equation}
i.e., the state with the minimum fluctuations \cite{Edmonds}. \ The spin
CS's are a generalization of the Holstein-Primakoff construction
\cite{Holstein}. \ They are defined in terms of deviations from the maximum
positive z-component of the spin $\mathbf{S}$
\begin{equation}
S_{z}|\mathbf{p}\rangle=\left(  j-p\right)  |\mathbf{p}\rangle \; .
\end{equation}
The CS's are then constructed using
\begin{eqnarray}
|\mathbf{\mu}\rangle &  =&\frac{1}{\left(  1+\left|  \mu\right|  ^{2}\right)  ^{j}}%
\exp\left(  \mu S_{-}\right)  |0\rangle\\
&  =&\frac{1}{\left(  1+\left|  \mu\right|  ^{2}\right)  ^{j}}\sum_{p=0}%
^{2j}\left(  \frac{\left(  2j\right)  !}{p!\left(  2j-p\right)  !}\right)
^{\frac{1}{2}}\mu^{p}|\mathbf{p}\rangle\nonumber
\end{eqnarray}

\noindent where $\mu$ is a complex number. \ Since the configuration space of
$\mathbf{S}$ is the surface of a sphere, it will be clearer to have $\mu$
parametrize the surface of a sphere through a stereographic projection,%

\begin{equation}
\mu=\tan\left(  \frac{1}{2}\theta\right)  e^{i\varphi}.
\end{equation}
In this representation, a CS will be represented  by a solid
angle $\mathbf{\Omega}$: \
\begin{equation}
|\mathbf{\Omega}\rangle=|\theta,\varphi\rangle=\left(  \cos\frac{1}{2}%
\theta\right)  ^{2j}\exp\left\{  \tan\left(  \frac{1}{2}\theta\right)
e^{i\varphi}S_{-}\right\}  |0\rangle \; .\label{def1}%
\end{equation}
A useful property for a path integral formulation is that the unit operator
has the familiar decomposition in terms of projection operators on all
CS's,
\begin{equation}
\frac{2j+1}{4\pi}\int d\mathbf{\Omega\quad}|\mathbf{\Omega}\rangle
\langle\mathbf{\Omega}|=1.
\end{equation}
In this representation, the overlap of two coherent states represents an area
on a sphere, the surface of which is 
the configuration space of the spin $\mathbf{S}$. \ The overlap is 
\begin{equation}
\langle\;\mathbf{\Omega}^{\prime}|\mathbf{\Omega\;}\rangle=\left\{  \cos
\frac{1}{2}\theta\cos\frac{1}{2}\theta^{\prime}+\sin\frac{1}{2}\theta\sin
\frac{1}{2}\theta^{\prime}e^{i\left(  \varphi-\varphi^{\prime}\right)
}\right\}  ^{2j}\label{def2}%
\end{equation}
and its magnitude is 
\begin{equation}
\left|  \langle\;\mathbf{\Omega}^{\prime}|\mathbf{\Omega\;}\rangle\right|
=\left(  \frac{1+\mathbf{n}\cdot\mathbf{n}^{\prime}}{2}\right)  ^{j} \; .%
\end{equation}

\bigskip

\noindent Since we plan to use a path-integral technique, we need to write
the expectation values of the Hamiltonian in the 
coherent representation. \ These expectation values follow in turn from 
those of the operators $S_{z}$, $S_{+}$ and 
$S_{-}$. \  The following expectation values 
 are deduced from Eq.$\left( \ref{def1}\right)$ and Eq.
$\left( \ref{def2}\right)$,
\begin{equation}
\langle\;\mathbf{\Omega}|j-S_{z}|\mathbf{\Omega\;}\rangle=j\left(
1-\cos\theta\right)  ,
\end{equation}%
\begin{equation}
\langle\;\mathbf{\Omega}|S_{+}|\mathbf{\Omega\;}\rangle=j\sin\theta
e^{i\theta},
\end{equation}%
\begin{equation}
\langle\;\mathbf{\Omega}|S_{-}|\mathbf{\Omega\;}\rangle=j\sin\theta
e^{-i\theta},
\end{equation}%
\begin{equation}
\langle\;\mathbf{\Omega}|\mathbf{S}|\mathbf{\Omega\;}\rangle=j\mathbf{n}.
\end{equation}
$\mathbf{n}$ is a unit vector with angles $\left(  \theta,\varphi\right)  $.
\ For $j\gg1$, the off-diagonal terms of the spin operator are smaller than the
diagonal ones by a factor of about $\sqrt{j}$. \ Hence they are
negligible in the classical limit. \ This limit will be implicit in all
subsequent calculations of the reduced density matrix elements.

\bigskip

\section{  REDUCED DENSITY MATRIX ELEMENTS OF THE SPIN PARTICLE    }

\bigskip

\indent In the following, we make use 
of CS for both the bath degrees of
freedom and the magnetic moment. \ The procedure we follow is by now 
mostly standard. Reference  \cite{Weiss} (and references therein) provides 
a general overview of these methods and hence we omit most of the 
intermediate steps in our calculation. 

\bigskip

  Bosonic CS's  were first used
 by Langer \cite{Langer} to study dissipation and fluctuations in a
superfluid, many-body problem. \ Starting from the equation of motion of the
density matrix of the whole system, $\rho$, a Landau-Ginzburg 
equation was recovered in the
equilibrium case 
and a Fokker-Planck equation in the classical limit,%

\begin{equation}
i \frac{\partial\rho}{\partial t}=\left[  \mathcal{H},\rho\right] \; .
\end{equation}
In a CS formulation, only diagonal elements of the reduced density matrix are
needed. \ Instead of starting from the above equation, we can instead start
from an integral representation of the density matrix elements. \ This method
is well known and is based on the Feynman-Vernon formalism \cite{Feynman}.
\ This path-integral approach is in real-time as opposed to the imaginary-time
approach in equilibrium thermodynamics. \ Hence questions like approach to
equilibrium can be studied within this approach. \ This method has seen many
different applications since the Caldeira-Leggett (CL) work \cite{Caldeira}.
The CL model was \ successful in showing how to recover the Langevin equation
by coupling an oscillator to a bath of oscillators. \ It seems natural then to
ask if the LLGB equation can be recovered by coupling a spin to a bath of
oscillators. \ This important question does not seem to have
been addressed in the literature. \ Spin coherent states are the
natural language to answer this question. \ Hence, we formulate 
the question in terms  of CS and use 
path-integral techniques to 
write the density matrix elements of the system. \ Use of path-integrals
with spin CS is not as straightforward as in the case of 
bosons  \cite{kochetov}, nevertheless it is the 
 method of choice  
in this particular problem. 

\bigskip

\indent      The calculation we present below takes into account the 
correct boundary
conditions as emphasized in (\cite{kochetov}). \ However 
we avoid using the more
abstract holomorphic representation in favor of a more geometric one, i.e., in
terms of solid angles. \ The physical space for the Hamiltonian , Eq.
$\left( \ref{hamiltonian}\right)$, is the product of the Hilbert space of 
the spin particle
and that of the harmonic oscillators,
\begin{equation}
\prod_{k}\left|  \mathbf{\Omega}\right\rangle \otimes\left|  \Phi
_{k}\right\rangle .
\end{equation}
Using the expectation values of the different operators in the Hamiltonian, we
get the expectation value of the Hamiltonian in the coherent representation,
\begin{eqnarray}
\mathcal{H}\left[  \Phi^{\ast},\Phi,\mathbf{S}\right]   &  = &-H_{z}j\cos\theta(t)+\sum
_{k}\omega_{k}\Phi_{k}^{\ast}(t)\Phi_{k}(t)\\
& & +j\sum_{k}\gamma_{k}\Phi_{k}^{\ast}(t)\sin\theta(t)e^{-i\varphi(t)}%
-j\sum_{k}\gamma_{k}^{\ast}\Phi_{k}(t)\sin\theta(t)e^{i\varphi(t)} . \nonumber
\end{eqnarray}
From now on, we normalize the magnitude of all spin vectors 
by $j$. \ The reduced density matrix element of the spin particle, 
$\rho_{ff'}$ , is by definition the
density matrix element of the whole system averaged over the states, 
$\left|\mathbf{\Phi}_{k}\right\rangle$, of the
bath,
\begin{eqnarray}
\rho_{ff^{\prime}}\left(  t\right)   &  =&\langle\mathbf{S}_{f^{\prime}%
}\;\left|  \rho(t)\right|  \;\mathbf{S}_{f}\rangle\\
&  =&\int\prod_{k}\mathfrak{D}\Phi_{k}^{\ast}\mathfrak{D}\Phi_{k}%
\;\langle\mathbf{S}_{f^{\prime}};\mathbf{\Phi}\;\left|  \rho(t)\right|  \;\mathbf{S}%
_{f};\mathbf{\Phi}\rangle.\nonumber
\end{eqnarray}
where $\left|\mathbf{S}_{f}\right\rangle$ and 
$\left|\mathbf{S}_{f'}\right\rangle$ are two arbitrary CS of the spin. 

  For simplicity, from now on we use the following notation for the functional
measure of the Bosonic degrees of freedom,
\begin{equation}
\int\prod_{k}\mathfrak{D}\Phi_{k}^{\ast}\mathfrak{D}\Phi_{k}\;\equiv
\;\int\mathfrak{D}\left(  \mathbf{\Phi}^{\ast},\mathbf{\Phi}\right)  .
\end{equation}

\bigskip

\indent The calculation of density matrix elements
 is easily carried out using a
path-integral representation. \ The propagator of the Bosonic part can be
written in terms of a path-integral \cite{Negele-Orland}
\begin{eqnarray}
\langle\mathbf{\Phi}_{f}\;|e^{-i\int_{0}^{t}dt\;\mathcal{H}\left(  t\right)
}|\;\mathbf{\Phi}_{i}\rangle &  =&\int_{\mathbf{\Phi}\left(  0\right)
=\Phi_{i}}^{\mathbf{\Phi}^{\ast}\left(  t\right)  =\Phi_{f}^{\ast}}%
\mathfrak{D}\left(  \mathbf{\Phi}^{\ast},\mathbf{\Phi}\right)  \;\exp\left\{
\sum_{k}\Phi_{k}^{\ast}\left(  t\right)  \Phi_{k}\left(  t\right)  \right.  \\
&  &\left.  +i\int_{0}^{t}dt\left[  \sum_{k}i\Phi_{k}^{\ast}\left(  t\right)
\partial_{t}\Phi_{k}\left(  t\right)  -\mathcal{H}\left(  \mathbf{\Phi}^{\ast
},\mathbf{\Phi,S}\right)  \right]  \right\}  .
\end{eqnarray}

\bigskip

Running from an initial time, $t=0$, to time $t$, we use a real-time path
integral to average over all intermediate states. \ The density matrix element
is then expressed as an integral in terms of the initial density matrix
element of the system,
\begin{eqnarray}
\rho_{ff^{\prime}}(t) &  =&\int\mathfrak{D}\left(  \Phi^{\ast},\Phi\right)
\int\mathfrak{D}\mathbf{\Omega}_{1}\int\mathfrak{D}\mathbf{\Omega}_{2}\int
\mathfrak{D}\left(  \Phi_{1}^{\ast},\Phi_{1}\right)  \int\mathfrak{D}\left(  \Phi
_{2}^{\ast},\Phi_{2}\right)  \label{density}\\
&  &\times\langle\mathbf{S}_{f},\Phi;t\;|\;\mathbf{\Omega}_{1},\Phi
_{1};0\rangle\langle\mathbf{\Omega}_{1},\Phi_{1};0\;|\rho|\;\mathbf{\Omega
}_{2},\Phi_{2};0\rangle\langle\mathbf{\Omega}_{2},\Phi_{2};0\;|\;\mathbf{S}%
_{f\prime},\Phi,t\rangle . \nonumber
\end{eqnarray}

\bigskip

We make no assumption about the initial state of the spin particle. \ Hence we
have to calculate a forward propagator, a backward propagator and the density
matrix element at the initial time. \ The system is assumed to be at finite
temperature. \ Since the Hamiltonian is quadratic, the integrations are easily
carried out in the stationary-phase approximation. \ We show a few steps 
in the 
 calculation of 
 the forward propagator. \ Similar calculations are also done for
the other two terms in Eq. $\left( \ref{density}\right)$. \ The forward 
propagator is first
written as a path integral
\begin{eqnarray}
\langle\mathbf{S}_{f},\Phi;t\;|\;\mathbf{\Omega}_{1},\Phi_{1};0\rangle &
=&\langle\;\mathbf{S}_{f},\Phi\left|  e^{-i\int_{0}^{t}\mathcal{H}dt}\right|
\mathbf{\Omega}_{1},\Phi_{1}\;\rangle\\
&  =&\int_{\mathbf{\Omega}_{1}}^{\mathbf{S}_{f}}\mathfrak{D}\mathbf{S}_{1}%
\int_{\Phi_{1}}^{\Phi}\mathfrak{D}\left(  \Phi_{1}^{\ast},\Phi_{1}\right)
\exp\left\{  \sum_{k}\Phi_{1,k}^{\ast}(t)\Phi_{1,k}(t)\right.  \nonumber\\
&  & \left.  +i\int_{0}^{t}dt^{\prime}\left[  i\sum_{k}\Phi_{1,k}^{\ast
}(t^{\prime})\partial_{t^{\prime}}\Phi_{1,k}(t^{\prime})-\mathcal{H}\left(
\Phi_{1}^{\ast},\Phi_{1},\mathbf{S}_{1}\right)  \right]  \right\}  \nonumber\\
& & \times\exp\left\{  i\mathbb{S}_{WZ}\left[  \mathbf{S}_{1}\right]  \right\}
.\nonumber
\end{eqnarray}
The last term is a geometrical term, the Wess-Zumino term
\cite{Fradkin}(and references therein)
\begin{equation}
\mathbb{S}_{WZ}\left[  \mathbf{S}_{1}\right]  =\int_{0}^{1}ds\int_{0}^{t}%
d\tau\mathbf{S}_{1}\left(  s,\tau\right)  \cdot\left(  \frac{\partial
\mathbf{S}_{1}\left(  s,\tau\right)  }{\partial s}\times\frac{\partial
\mathbf{S}_{1}\left(  s,\tau\right)  }{\partial\tau}\right)  \; ,
\end{equation}
where $\mathbf{S}_{1}\left(s,\tau \right)$ is a  homotopy map 
between the side $\left( \mathbf{z}, \mathbf{\Omega}_{1} \right)$ and 
the side $\left( \mathbf{z}, \mathbf{S}_{f} \right)$ \cite{Whitehead}.
This term therefore 
represents the area enclosed by the trajectory of the spin vector, and
hence there is a corresponding two-form \cite{Permov}. Using 
Stoke's theorem, it can be
written in terms of a path integral. It is also well known
\cite{Permov} that in this form the 1-form that results is physically the
potential of a magnetic monopole at the center of a sphere. \ Since 
 a full discussion of the topological nature of this term is outside
the scope of this paper,  
we refer to the above literature for further details. \ The bath degrees of
freedom are eliminated by a stationary-phase evaluation of the integral. \ The
phase is an extremum for states that satisfy
\begin{equation}
i\partial _{t}{\Phi}_{1,k}\left( t \right)=\frac{\delta\mathcal{H}}{\delta\Phi_{1,k}^{\ast}(t)}
\end{equation}
and a similar equation for $\Phi^{*}_{1,k}$. \ Summations over 
k are implicit in
what follows. \ We have
\begin{equation}
i\partial _{\tau}{\Phi}_{1,k}(\tau)=\omega_{k}\Phi_{1,k}(\tau)-\gamma_{k}%
(\tau)S_{-}%
\end{equation}%
\begin{equation}
i\partial_{\tau}{\Phi}_{1,k}^{\ast}(\tau)=\omega_{k}\Phi_{1,k}^{\ast}%
(\tau)-\gamma_{k}^{\ast}(\tau)S_{+}.
\end{equation}
The solutions with the correct boundary conditions are 
\begin{eqnarray}
\Phi_{i,k}(\tau)&=&\Phi_{1,k}e^{-i\omega_{k}\tau}+ie^{-i\omega_{k}\tau}\int
_{0}^{\tau}dt^{\prime}e^{i\omega_{k} t^{\prime}}\gamma_{k}(t^{\prime}%
)S_{-}(t^{\prime})\\
\Phi_{i,k}^{\ast}(\tau)&=&\Phi_{k}^{\ast}e^{i\omega_{k}(\tau-t)}+ie^{i\omega
_{k}\tau}\int_{\tau}^{t}dt^{\prime}e^{-i\omega_{k}t^{\prime}}\gamma_{k}^{\ast
}(t^{\prime})S_{+}(t^{\prime}) \nonumber \\
& &0\leq\tau\leq t \nonumber.
\end{eqnarray}
At the endpoints, we then have:
\begin{eqnarray}
\Phi_{1,k}(t)&=&\Phi_{1,k}e^{-i\omega_{k}t}+ie^{-i\omega_{k}t}\int_{0}%
^{t}dt^{\prime}e^{i\omega_{k} t^{\prime}}\gamma_{k}(t^{\prime}%
)S_{-}(t^{\prime})\\
\Phi_{1,k}^{\ast}(0)&=&\Phi_{k}^{\ast}e^{-i\omega_{k}t}+i\int_{0}^{t}%
dt^{\prime}e^{-i\omega_{k}t^{\prime}}\gamma_{k}^{\ast}(t^{\prime}%
)S_{+}(t^{\prime}).\nonumber
\end{eqnarray}

\bigskip

These solutions are then put back in Eq. $\left(\ref{density}\right)$. \ Similar
expressions follow from the calculations of the backward propagator. \ The
density matrix element at the initial time is calculated with the assumption
that the bath is initially at equilibrium with the spin. \ The bath relaxes
much faster than the spin, a reasonable approximation in many  problems
in magnetics. \ In this case the density matrix is separable at the initial
time. \ The bath density matrix, $\rho_{B}$, is 
then known and its matrix elements can be written explicitly in terms 
of the Hamiltonian, $\mathcal{H}_{B}$, of the bath only,
\begin{eqnarray}
\langle\;\Phi_{1}\left|  \rho_{B}(0)\right|  \Phi_{2}\;\rangle &  =&\frac
{1}{Z_{B}}\langle\Phi_{1}\left|  e^{-\beta \mathcal{H}_{B}}\right|  \Phi_{2}\rangle\\
&  &=\frac{1}{Z_{B}}\int_{\Phi_{2}}^{\Phi_{1}}\mathfrak{D}\Phi\exp\left\{
\;\Phi^{\ast}(\beta)\Phi(\beta)\right.  \nonumber\\
& & +i\int_{0}^{\beta}d\tau\left.  \left[  i\Phi^{\ast}(\tau)\Phi
(\tau)-\mathcal{H}(\tau)\right]  \;\right\}  \nonumber
\end{eqnarray}
with periodic boundary conditions
\begin{eqnarray}
\mathbf{\Phi}(0) &  = &\mathbf{\Phi}_{2}\\
\mathbf{\Phi}^{\ast}(\beta) &  =&\mathbf{\Phi}_{2}^{\ast} . \nonumber
\end{eqnarray}
We find, after applying a stationary-phase approximation to the integral, the
expression
\begin{equation}
\langle\Phi_{1}\left|  \rho_{B}(0)\right|  \Phi_{2}\rangle=\exp\left\{
\sum_{k}\Phi_{1,k}^{\ast}\Phi_{2,k}e^{-\beta\omega_{k}}\right\}  .
\end{equation}
After integrating out the degrees of freedom of the bath, we are left 
 with only
integrals over paths in the spin configuration space. \ The effective action
of the spin is now complex, as usual with dissipative systems. \ The
reduced density matrix element is now given by
\begin{eqnarray}
\rho_{ff^{\prime}}\left(  t\right)   &  =&\int\mathfrak{D}\mathbf{\Omega}_{1}%
\int\mathfrak{D}\mathbf{\Omega}_{2}\;\langle\mathbf{\Omega}_{1}\left|  \rho
_{s}\left(  0\right)  \right|  \mathbf{\Omega}_{2}\rangle\label{reduced}\\
& & \times
\int_{\mathbf{\Omega}_{1}}^{\mathbf{S}_{f}}\mathfrak{D}\mathbf{S}_{1}%
\int_{\mathbf{S}_{f^{\prime}}}^{\mathbf{\Omega}_{2}}\mathfrak{D}\mathbf{S}_{2}\exp\left\{
iH_{z}\int_{0}^{t}dt^{\prime}\left(  S_{1,z}\left(  t^{\prime}\right)
-S_{2,z}\left(  t^{\prime}\right)  \right)  \right.  \nonumber\\
& & \left.  +iS_{WZ}\left[  \mathbf{S}_{1}\right]  -iS_{WZ}\left[
\mathbf{S}_{2}\right]  \;\right\}  \times\mathcal{W}\left(  \mathbf{S}%
_{1},\mathbf{S}_{2}\right)  \nonumber
\end{eqnarray}
where the last term is entirely due to the coupling between the bath and the
spin particle. It is given by
\begin{eqnarray}
\ln\mathcal{W}\left(  \mathbf{S}_{1},\mathbf{S}_{2}\right)   &  =-
\int_{0}^{t}dt^{\prime}\int_{0}^{t}dt^{\prime\prime}\;\exp\left(  -i\omega
_{k}\left(  t^{\prime}-t^{\prime\prime}\right)  \right)  \gamma_{k}^{\ast
}\left(  t^{\prime}\right)  \gamma_{k}\left(  t^{\prime\prime}\right)  \\
&  \times\left\{  \;\Theta\left(  t^{\prime\prime}-t^{\prime}\right)
S_{2,+}\left(  t^{\prime}\right)  S_{2,-}\left(  t^{\prime\prime}\right)
\right.  \nonumber\\
&  \left.  +\left(  1-\Theta\left(  t^{\prime\prime}-t^{\prime}\right)
\right)  S_{1,+}\left(  t^{\prime}\right)  S_{1,-}\left(  t^{\prime\prime
}\right)  \;\right\}  \nonumber\\
&  +\int_{0}^{t}dt^{\prime}\int_{0}^{t}dt^{\prime\prime}\;\exp\left(
-i\omega_{k}\left(  t^{\prime}-t^{\prime\prime}\right)  \right)  \gamma
_{k}^{\ast}\left(  t^{\prime}\right)  \gamma_{k}\left(  t^{\prime\prime
}\right)  \nonumber\\
&  \qquad S_{2,+}\left(  t^{\prime}\right)  S_{1,-}\left(  t^{\prime\prime
}\right)  \nonumber\\
&  +\frac{1}{e^{\beta\omega_{k}}-1}\int_{0}^{t}dt^{\prime}\int_{0}%
^{t}dt^{\prime\prime}\exp\left(  -i\omega_{k}\left(  t^{\prime}-t^{\prime
\prime}\right)  \right)  \gamma_{k}^{\ast}\left(  t^{\prime}\right)
\gamma_{k}\left(  t^{\prime\prime}\right)  \nonumber\\
&  \times\left\{  \;S_{1,+}\left(  t^{\prime}\right)  S_{2,-}\left(
t^{\prime\prime}\right)  +S_{2,+}\left(  t^{\prime}\right)  S_{1,-}\left(
t^{\prime\prime}\right)  \right\}  \nonumber\\
&  -\frac{1}{e^{\beta\omega_{k}}-1}\int_{0}^{t}dt^{\prime}\int_{0}%
^{t}dt^{\prime\prime}\exp\left(  i\omega_{k}\left(  t^{\prime}-t^{\prime
\prime}\right)  \right)  \gamma_{k}\left(  t^{\prime}\right)  \gamma_{k}%
^{\ast}\left(  t^{\prime\prime}\right)  \nonumber\\
&  \times\left\{  \;S_{1,+}\left(  t^{\prime\prime}\right)  S_{1,-}\left(
t^{\prime}\right)  +S_{2,+}\left(  t^{\prime\prime}\right)  S_{2,-}\left(
t^{\prime}\right)  \right\} . \nonumber
\end{eqnarray}

\bigskip

By taking the limit of an infinite number of oscillators, this latter term
becomes responsible for the appearance of dissipation in this model. \ After
calculating the elements of the reduced density matrix, we can now calculate
its time evolution and find a Fokker-Plank type equation as was done in the
original work of CL \cite{Caldeira}. \ We choose rather to take the
semiclassical limit of this expression and see under what conditions, if any,
a LLGB equation can be recovered. \ 

\bigskip

\bigskip

\section{THE SEMI-CLASSICAL APPROXIMATION  }

\bigskip 

In this section, we find the equation of motion of the magnetization
by calculating the most probable configurational paths. \ This is done
 by calculating the path in the reduced density matrix element for 
the spin field that has
the largest weight. \ Then we show that these paths are really the 
semiclassical limit of the classical paths averaged over the thermal 
fluctuations in the LLGB equation. \ We also show that the 
fluctuation-dissipation theorem is satisfied. It reduces to the Brown form 
only in the high temperature limit and only in the linear response
approximation. \ This approximation  fails when the system is highly 
anisotropic \cite{bertram2}.

\bigskip

\indent To facilitate the taking of  the classical limit we make the
 change of variables,
\begin{eqnarray}
\mathbf{S}\left(  \tau \right)   & = \frac{1}{2}\left(  \mathbf{S}_{1}\left(
\tau \right)  +\mathbf{S}_{2}\left(  \tau \right)  \right)  ,\\
\mathbf{D}\left(  \tau \right)   & = \mathbf{S}_{1}\left(  \tau \right)
-\mathbf{S}_{2}\left(  \tau \right)  .\nonumber
\end{eqnarray}
The variable $\mathbf{D}$ represents the fluctuating field that is coupled to
the spin and is due to the inherent irreversibility in the system. In terms of
these new variables, the weight functional $\mathcal{W}\left(  \mathbf{S}%
_{1},\mathbf{S}_{2}\right)  $ becomes
\begin{eqnarray}
\mathcal{W}\left[  \mathbf{S},\mathbf{D}\right]   &  =\exp\left\{  
\int_{0}^{t}dt^{\prime}\int_{0}^{t}dt^{\prime\prime}e^{-i\omega
_{k}t^{\prime}}e^{i\omega_{k}t^{\prime\prime}}\gamma_{k}^{\ast}\left(
t^{\prime}\right)  \gamma_{k}\left(  t^{\prime\prime}\right)  \right. \\
&  \times\left\{  -\frac{1}{2}D_{+}\left(  t^{\prime}\right)  D_{-}\left(
t^{\prime\prime}\right)  +\Theta\left(  t^{\prime\prime}-t^{\prime}\right)
S_{+}\left(  t^{\prime}\right)  D_{-}\left(  t^{\prime\prime}\right)  \right.
\nonumber\\
&  -\left.  \left(  1-\Theta\left(  t^{\prime\prime}-t^{\prime}\right)
\right)  D_{+}\left(  t^{\prime}\right)  S_{-}\left(  t^{\prime\prime}\right)
\right\} \nonumber\\
&  -\left.  \frac{1}{e^{\beta\omega_{k}}-1}\int_{0}^{t}dt^{\prime}\int
_{0}^{t}dt^{\prime\prime}e^{-i\omega_{k}t^{\prime}}e^{i\omega
_{k}t^{\prime\prime}}\left(  D_{+}\left(  t^{\prime}\right)  D_{-}\left(
t^{\prime\prime}\right)  \right)  \right\} \nonumber
\end{eqnarray}
where $\Theta \left( \tau \right)$ is the unit step function
\begin{equation}
\Theta \left( \tau \right) = \frac{1}{2} + \frac{1}{2\pi i} \mathcal{P} \int_{-\infty}
^{\infty}  e^{i\omega \tau}\; \frac{d\omega}{\omega}.
\end{equation}

\bigskip

\noindent In these new variables, the phase $ln \mathcal{W}$ clearly shows that 
the action of the reservoir results in an extra linear coupling in 
$\mathbf{S}$ and $\mathbf{D}$. \ Moreover, we now have  
a quadratic term involving the variable $\mathbf{D}$. \ This quadratic term
is easily seen to be real and negative, assuring convergence of the 
sum over all configurations of $\mathbf{D}$. \ The linear
term describing interaction of 
 the fields $\mathbf{S}$ and $\mathbf{D}$ is  
imaginary, however.  \ In fact, it is such a 
term that gives rise to dissipation
in the energy of the spin $\mathbf{S}$. \ As we will show below,  
 $\mathbf{D}$ is the field that is associated with the classical 
random field in the LLGB equation. \ If the coupling 
term in $\ln\mathcal{W}$ is  set to
zero, then integrations over $\mathbf{D}$ are equivalent to averaging
over a Gaussian fluctuating field. \ As will be seen below, this is the limit
in which we recover the LLGB equation. \ In this case the bath 
does not depend on $\mathbf{S}$. \ This approximation fails when the 
Hamiltonian is not symmetric under the interchange of the dynamical
spin components.

To find the semiclassical result for the reduced density matrix, we again
resort to a stationary phase approximation to the phase of the path integrals
in Eq.$\left(\ref{reduced} \right)$. \ First we impose  constraints  on the spin
magnitude by introducing two Lagrangian multiplier fields, $\eta_{1}\left(  t\right)  $
and $\eta_{2}\left( t\right)  $. \ These are then coupled to the spin fields 
$\mathbf{S}_{1}$ and $\mathbf{S}_{2}$, respectively, 
as follows,%

\begin{equation}
\delta\left(  \mathbf{S}_{1}^{2}\left(  \tau \right)  -1\right)  =\int\mathfrak{D}%
\eta_{1}\left(  \tau \right)  \;e^{i\int d \tau\;\eta_{1}\left(  \tau \right)  \left(
\mathbf{S}_{1}^{2}\left(  \tau \right)  -1\right)  }\label{constraint}.
\end{equation}
A similar expression holds for the other spin vector 
variable, $\mathbf{S}_{2}$, and $\eta_{2}$. \ These 
constraints are then put back in the 
expression for the reduced density matrix element. \ The phase of the 
path integral is now a function of four independent fields 
$\eta_{1}$, $\eta_{2}$, $\mathbf{S}$ and $\mathbf{D}$. \ Extremizing
the phase of these paths gives the semiclassical solution in the large
spin limit. \
After solving for the constrained fields in terms of the spin fields 
$\mathbf{S}_{1}$ and $\mathbf{S}_{2}$, we write the remaining two equations
in $\mathbf{S}$ and $\mathbf{D}$ only, obtaining 
a generalized   
 form of the LLGB equation

\begin{equation}
\frac{d\mathbf{S}\left(  t^{\prime}\right)  }{d t^{\prime}%
}=\mathbf{S\times}\left(  \mathbf{H}+\mathbf{T}^{\left(  S\right)
}+\mathbf{T}^{\left(  D\right)  }\right)  +\frac{1}{4}\mathbf{D}%
\times\mathbf{W} \; , \label{llg1}
\end{equation}%
and

\begin{equation}
\frac{d \mathbf{D}\left(  t^{\prime}\right)  }{d t^{\prime}%
}=\mathbf{D}\times\left(  \mathbf{H}+\mathbf{T}^{\left(  S\right)  }\right)
+\mathbf{S}\times\mathbf{W}+\mathbf{D}\times\mathbf{T}^{\left(  D\right)  } . \label{llg2}
\end{equation}

\bigskip 

 The vectors $\mathbf{T}^{\left(  S\right)  }$, 
$\mathbf{T}^{\left(  D\right)  }$ and $\mathbf{W}$ are associated with
dissipation, thermal fluctuations and magnitude fluctuations, 
respectively. \ In Cartesian 
form,  they are
respectively given in terms of the functions $J$ and $F_{\beta}$ by

\begin{equation}
\mathbf{T}^{\left(  S\right)  }\left(  u\right)  =\left[
\begin{array}
[c]{c}%
i\int_{0}^{u}dt^{\prime}\Theta\left(  u-t^{\prime}\right)  \left\{
J\left(  u-t^{\prime}\right)  S_{-}\left(  t^{\prime}\right)  -J^{\ast}\left(
u-t^{\prime}\right)  S_{+}\left(  t^{\prime}\right)  \right\}  \\
\\
-\int_{0}^{u}dt^{\prime}\Theta\left(  u-t^{\prime}\right)  \left\{
J\left(  u-t^{\prime}\right)  S_{-}\left(  t^{\prime}\right)  +J^{\ast}\left(
u-t^{\prime}\right)  S_{+}\left(  t^{\prime}\right)  \right\}  \\
\\
0
\end{array}
\right],
\end{equation}

\bigskip%

\begin{equation}
\mathbf{T}^{\left(  D\right)  }\left(  u\right)  =\left[
\begin{array}
[c]{c}%
{i}\int_{0}^{t}dt^{\prime}\left\{  \left(  \frac{1}{2}J\left(  u-t^{\prime
}\right)  +F_{\beta}\left(  u-t^{\prime}\right)  \right)  D_{-}\left(
t^{\prime}\right)  \right.  \\
\left.  +\left(  \frac{1}{2}J^{\ast}\left(  u-t^{\prime}\right)  +F_{\beta
}^{\ast}\left(  u-t^{\prime}\right)  \right)  D_{+}\left(  t^{\prime}\right)
\right\}  \\
\\
-\int_{0}^{t}dt^{\prime}\left\{  \left(  \frac{1}{2}J\left(  u-t^{\prime
}\right)  +F_{\beta}\left(  u-t^{\prime}\right)  \right)  D_{-}\left(
t^{\prime}\right)  \right.  \\
\left.  -\left(  \frac{1}{2}J^{\ast}\left(  u-t^{\prime}\right)  +F_{\beta
}^{\ast}\left(  u-t^{\prime}\right)  \right)  D_{+}\left(  t^{\prime}\right)
\right\}  \\
\\
0
\end{array}
\right],
\end{equation}
and
\bigskip

\bigskip%

\begin{equation}
\mathbf{W}\left(  u\right)  =\left[
\begin{array}
[c]{c}%
{-i}\int_{0}^{t}dt^{\prime}\Theta\left( t^{\prime}-u \right)  \left\{
\;J\left(  u-t^{\prime}\right)  D_{-}\left(  t^{\prime}\right)  -J^{\ast
}\left(  u-t^{\prime}\right)  D_{+}\left(  t^{\prime}\right)  \right\}  \\
\\
\int_{0}^{t}dt^{\prime}\Theta\left( t^{\prime}-u\right)  \left\{
\;J\left(  u-t^{\prime}\right)  D_{-}\left(  t^{\prime}\right)  +J^{\ast
}\left(  u-t^{\prime}\right)  D_{+}\left(  t^{\prime}\right)  \right\}  \\
\\
0
\end{array}
\right].
\end{equation}

\bigskip

\noindent These vectors are in general non-local in time and hence 
include memory effects in the equations of motion for $\mathbf{S}$ and 
$\mathbf{D}$. \ This type of behavior is clearly needed when the relaxation
time of the reservoir is of the same order as that of the spin particle. \ We 
will not discuss such a situation here. \ We are mainly interested to recover 
the constant dissipation case. \ Even though we called Eq.$\left(
 \ref{llg1}\right)$ and Eq.$\left( \ref{llg2} \right)$
generalized LLGB equations, it is {\textit{not}} yet clear 
 how a Gilbert damping
term can arise in these equations. \ However through a careful choice of 
the density of states of the bath, the coupling constants and the initial 
conditions, such damping form can be recovered as shown below. 

\bigskip

\noindent To describe dissipation, we take the 
continuum limit for the bath states. Then the spectral functions
$J$ and $F_{\beta}$ are given by

\begin{equation}
J\left(  \tau-\tau^{\prime}\right)  =\int_{0}^{\infty} d\omega \; 
\lambda\left(  \omega\right)
\left|  \gamma\left(  \omega\right)  \right|  ^{2}\exp\left[  -i\omega\left(
\tau-\tau^{\prime}\right)  \right] \label{J}
\end{equation}
and
\bigskip%

\begin{equation}
F_{\beta}\left(  \tau-\tau^{\prime}\right)  =\int_{0}^{\infty} d\omega \; 
\frac{\lambda\left(
\omega\right)  }{\exp\left[  \beta\omega\right]  -1}\left|  \gamma\left(
\omega\right)  \right|  ^{2}\exp\left[  -i\omega\left(  \tau-
\tau^{\prime}\right)
\right] . \label{Fb}
\end{equation}

\bigskip

\noindent $F_{\beta}$ is simply 
the nonzero temperature counterpart of $J$. \  
$\lambda \left( \omega \right)$ is the density of states of the bath. \ In 
fact the function,
\begin{equation}
\mathcal{G} \left( \tau^{\prime}  - \tau \right) = \frac{1}{2}J\left(  \tau^{\prime}
- \tau\right)  +F_{\beta}\left(  \tau^{\prime
}- \tau\right),
\end{equation}
is the inverse of the free propagator of the field $\mathbf{D}$.

\bigskip 

 The vectors $\mathbf{S}$ and $\mathbf{D}$ are 
orthogonal as follows
from the constraint equations, Eq.$\left( \ref{constraint}\right)$. \ Note that 
when $\mathbf{D}$ is set to zero,
the density matrix  becomes diagonal but 
the equation of motion for $\mathbf{S}$ will still have an extra term
besides the precessional term that is 
due to the external field $\mathbf{H}$. \ This extra 
term $\mathbf{T}^{\left(  S\right)  }\left(  u\right)$ 
clearly 
always has a damping effect. \ We conclude that it is the vector
$\mathbf{S}$ that must be identified with the classical 
magnetization and that  
$\mathbf{D}$  is the part that gives rise to the thermal fluctuations 
in $\mathbf{S}$. \ Finally, we observe that the last term 
in Eq.$\left(\ref{llg1}\right)$ can not be recovered in the classical limit. \ This 
quantum mechanical term is  not 
present in the LLGB equation and is beyond a classical 
linear-response treatment
of the problem of fluctuations. \ It is of higher order
in $\mathbf{D}$ and temperature-independent. \ It is easy 
to see that this term 
gives rise to fluctuations in the magnitude 
of the spin. \ These fluctuations can not be 
accounted for classically since the magnitude of the magnetization is assumed 
to be constant.

 One important thing to note from Eq.$\left(\ref{llg1}\right)$ is that
the vector $\mathbf{T}^{\left(  D\right)  }$ is 
complex and hence, if fluctuations are present, the equation of motion for
$\mathbf{S}$ becomes complex. \ The physical interpretation 
of this equation then becomes obscure at this level and may not be 
used as it stands to get the effective magnetization of the particle. \ Having
a complex equation for the extremum path of the spin is however expected
given that the same result happens in the case of the harmonic 
oscillator \cite{Caldeira}. \ A solution for the fluctuating 
magnetization is then sought through a 
direct calculation of the propagators in Eq.$\left( \ref{density}\right)$.

\bigskip

\indent Now we  show that a generalized fluctuation-dissipation
theorem is satisfied as expected for this system since 
we started from a closed system and integrated  out a large 
part of its degrees of freedom. \ We will also 
show that it is the vector $\mathbf{D}$ that should be regarded 
as the quantum source of the thermal fluctuations in the 
spin system as treated in LLGB. \ The vector $\mathbf{S}$ is then 
the physically measurable magnetization. \ To better 
understand the physical meaning of the field $\mathbf{D}$ and to 
recover the standard stochastic description 
of the thermal fluctuations, we introduce
yet another field, $\mathbf{\xi} \left( t \right)$. \ In Eq.$\left( \ref{density}\right)$, 
we will replace the l.h.s. of the following expression
with the r.h.s. 

\begin{eqnarray}
 exp\left( - \int_{0}^{t}dt^{\prime} \int_{0}^{t}dt^{\prime\prime} \mathbf{D}
_{+}\left(t^{\prime}\right) \mathcal{G} \left( t^{\prime}-t^{\prime\prime}\right)
\mathbf{D}_{-} \left( t^{\prime\prime}\right) \right) =
\end{eqnarray}
\begin{equation*} 
 \mathcal{N}
\int \mathfrak{D} \xi \; exp \left( -\frac{1}{2} 
\int_{0}^{t}dt^{\prime} \int_{0}^{t}dt^{\prime\prime} \xi_{l} \left( t^{\prime}
\right) \frac{1}{2} \mathcal{G}_{ll}^{-1}\left( t^{\prime}-t^{\prime\prime}\right)
\xi_{l} \left( t^{\prime \prime} \right) - i \int_{0}^{t} dt^{\prime}\xi_{l}
\left( t^{\prime} \right)\mathbf{D}_{l}
\left( t^{\prime} \right) \right) 
\end{equation*}
where $\mathcal{N}$ is a normalization constant.  \ The 
 quadratic 
term in $\mathbf{D}$  is then assumed to be a result of an averaging 
over all configurations of the field $\mathbf{\xi}$. \ Hence the 
path-integral for the reduced density, Eq.$\left( \ref{density}\right)$, is now 
in terms of three fields. \ The field $\mathbf{\xi}$ will now result
in a third  
 equation of motion.  
\ However to recover a thermal field similar to that introduced 
by Brown \cite{brown}, we proceed by 
assuming that the field $\mathbf{\xi}$ is  
 classical, i.e., we ignore its equation of motion. \ At this point
 the fields $\mathbf{\xi}$ and $\mathbf{H}$ are treated as 
non-dynamical fields. \ Next we solve for the magnetization
$\mathbf{S}$ for a given $\mathbf{\xi}$ and only then do we average over all 
configurations of $\mathbf{\xi}$ with the quadratic weight that 
we originally ignored in the solution. \ In fact 
$\mathbf{\xi}$ becomes the Brown stochastic 
field if we take the classical limit, that of high temperature. \ Given 
these observations, we can now assume that the effective 
thermal field with which the spin is interacting is 
really nothing more than the abstract $\mathbf{D}$ field that has 
been introduced in this calculation of the reduced
density matrix element. \ In fact  $\mathbf{\xi}$ has the following
correlation functions

\begin{eqnarray}
\langle \xi_{l}\left(  \tau\right)  \xi_{l^{\prime}}\left(  \tau^{\prime}\right)  \rangle &
= & 2 \delta_{ll^{\prime}} \mathcal{G}\left( \tau- \tau^{\prime}\right)  \\
&  =&\frac{1}{\pi}\delta_{ll^{\prime}}\int_{0}^{\infty} d\omega\;{\omega}
\coth\left(  \frac{\beta\omega
}{2}\right)  \frac{\pi\lambda\left(  \omega\right)  \left|  \gamma\left(
\omega\right)  \right|  ^{2}}{\omega}\exp\left[  -i\omega\left(  \tau
-\tau^{\prime
}\right)  \right] . \nonumber
\end{eqnarray}

\bigskip

\indent To recover the correlations of 
the thermal field assumed in the LLGB equation, we simply take 
the high temperature limit and require that the bath satisfies the 
 condition,
\begin{equation}
\frac{\pi\lambda\left(  \omega\right)  \left|  \gamma\left(  \omega\right)
\right|  ^{2}}{\omega}=\alpha,\label{Condition}
\end{equation}
where $\alpha$ is a constant. \ This condition provides the simplest
relation between fluctuations and dissipation. \ In this case, the 
correlation functions
for the random field become simply
\begin{equation}
\langle \xi_{l}\left(  \tau\right)  \xi_{l^{\prime}}\left(  \tau^{\prime}\right)  \rangle=2 \delta_{ll^{\prime}}\alpha
kT\delta\left(  \tau^{\prime}-\tau\right) .
\end{equation}
A similar condition arises if we replace the spin degrees of freedom by 
those of a 
 harmonic oscillator \cite{Caldeira}. \ However at 
high temperature, as we noticed earlier, 
a large spin can  be approximated well by an oscillator. \ This condition 
is, however, still true even if 
the bosonic degrees of freedom of the bath are replaced 
by fermionic degrees of freedom. 
\bigskip

\indent Finally we consider recovering the LLG 
equation with the Gilbert form of damping. \ Equations $\left( \ref{llg1}
\right)$ and 
$\left( \ref{llg2} \right)$ are very
general as they stand and it is not clear if the dissipation has 
the Gilbert form. \ To deduce the very special case of 
constant damping with the Gilbert form,    
 we set 
the fluctuations to zero and take
the following form for the spectral function $J$,
\begin{equation}
J\left(  \tau^{\prime}-\tau\right)  =i\alpha\frac{d}{dt}\delta
\left(  \tau^{\prime
}-\tau\right) . \label{Jeq}
\end{equation}
After an integration by parts of the term containing $\mathbf{T}^{\left( 
S \right)}$ in 
 the reduced density matrix element, Eq.$\left(\ref{density}\right)$, the equation of motion for $\mathbf{S}$, the magnetization, 
becomes simply 
\begin{eqnarray}
\frac{d\mathbf{S}\left(  \tau\right)  }{d \tau}=\mathbf{S}\left(
\tau\right)  \times \left(     \mathbf{H} +   \alpha \left(
\frac{d\mathbf{S} \left( \tau \right) }{d \tau} - \frac{d \; \mathbf{S}\cdot
\mathbf{z}}{d \tau} \mathbf{z} \right)         \right)
\end{eqnarray}


The boundary terms in the integration by parts are easily dealt with by 
a renormalization of the measure of the path-integral. \ Keeping in mind 
the model used to derive this result, this equation 
reduces to the LLG form only in the limit of small deviations from local
equilibrium. \
It is also important to note
that the choice we made for $J$, Eq.$\left(\ref{Jeq}\right)$,
 is compatible with the Gaussian 
approximation for the thermal field. \ Hence the stochastic LLG equation
is compatible with the FDT at high 
temperatures. \ However in this case, the application of the FDT to the LLG 
equation is not so trivial as for the generalized LLGB equations, Eqs. 
$\left( \ref{llg1}\right),\left(\ref{llg2}\right)$. \ In this case the correlation functions will depend 
on the dynamics of the system, i.e., the symmetry of the Hamiltonian.

\bigskip

 Before we end this section, we make a final comment about
the condition, Eq.$\left( \ref{Condition}\right)$, by which we recovered the LLG 
limit. \ If we 
assume constant coupling constants for the interaction between the 
spin and the bath, we find that the density of states must be linear. \ For 
phonons, the density of states is quadratic and hence, based on this 
assumption, can not  be the major source of the damping
constant $\alpha$. \ In fact dissipation due to currents is believed 
to be much larger \cite{Over}. \ Ferromagnetic 
compounds, such as FeNi, show a complex 
density of states for the non-localised electrons, hence a condition 
such as that given in Eq.$\left( \ref{Condition}\right)$ is 
representative of many competing 
mechanisms.\ It is only 
 the lower part of the spectrum that is important for a constant 
dissipation. \ In fact in Eq.$\left( \ref{J}\right)$, the limit of 
integration can not 
be taken to be infinite for a real bath. \ This in turn will introduce 
a new cut-off parameter in condition Eq.$\left( \ref{Condition}\right)$ which will 
be system dependent.  
 
\bigskip

\section{\bigskip ANISOTROPIC PARTICLES   }

\bigskip

 \ Perpendicular recording 
requires particles with 
relatively high anisotropy for long-term storage purposes. \ Hence 
a relevant question to
ask is how does anisotropy interact with the thermal field. \ This requires a
treatment beyond the linear response approach. \ We treat this question in
this last section. \ We limit ourselves to the simplest case; that of uniaxial
anisotropy along the z-axis. \ In this case 
the Hamiltonian of the particle-bath
system becomes%

\begin{equation}
\mathcal{H}=-H_{z}S_{z}-KS_{z}^{2}+\sum\omega_{k}a_{k}^{+}a_{k}-\sum\gamma
_{k}a_{k}^{+}S_{-}-\gamma_{k}^{\ast}S_{-}a_{k},
\end{equation}
where $K$ is the anisotropy constant. \ The external field is 
taken along the easy axis. \ Similarly to the 
above calculation, we find that the reduced
density matrix elements in the presence of anisotropy become%

\begin{eqnarray}
\rho_{ff^{\prime}}\left(  t\right)   &  =&\int\mathfrak{D}\mathbf{\Omega}%
_{1}\int\mathfrak{D}\mathbf{\Omega}_{2}\;\langle\;\mathbf{\Omega}_{1}\left|
\rho_{s}\left(  0\right)  \right|  \mathbf{\Omega}_{2}\;\rangle
\label{anisotropy}\\
&  &\times\int_{\mathbf{\Omega}_{1}}^{\mathbf{S}_{f}}\mathfrak{D}\mathbf{S}%
_{1}\int_{\mathbf{S}_{f^{\prime}}}^{\mathbf{\Omega}_{2}}\mathfrak{D}\mathbf{S}_{2}%
\exp\left\{  iH_{z}\int_{0}^{t}dt^{\prime}\left(  \;S_{1,z}\left(  t^{\prime
}\right)  -S_{2,z}\left(  t^{\prime}\right)  \right)  \right.  \nonumber\\
& & \left.  +iK\int_{0}^{t}dt^{\prime}\left(  S_{1,z}^{2}\left(
t^{\prime}\right)  -S_{2,z}^{2}\left(  t^{\prime}\right)  \right)  +i\left(
\;S_{WZ}\left[  \mathbf{S}_{1}\right]  - S_{WZ}\left[  \mathbf{S}_{2}\right]
\right)  \right\}  \mathcal{W}\left(  \mathbf{S}_{1},\mathbf{S}_{2}\right) .
\nonumber 
\end{eqnarray}

\bigskip

\noindent Clearly the bath influence on the magnetic moment  with 
and without anisotropy is the same as before, but now there is 
coupling between
the fluctuating field and the spin field that is anisotropy-dependent.
\ Therefore the random field becomes $K$ -dependent beyond the linear-response
approximation. \ We show below how to calculate the new correlation functions
of the fluctuations beyond the Gaussian approximation. \ We define two new
anistropy related vectors,%

\begin{eqnarray}
\mathbf{K}_{S}  &  = & \widehat{z}\left(  2K  S_{z}\right) \; ,
\end{eqnarray}
and
\begin{eqnarray} 
\mathbf{K}_{D}  &  = & \widehat{z}\left(  2K  D_{z}\right) .
\end{eqnarray}
The equations of motion for the spin field and the fluctuating field become 
in this case 

\begin{equation}
\frac{d\mathbf{S}\left(  t^{\prime}\right)  }{d t^{\prime}%
}=\mathbf{S\times}\left(  \mathbf{H}+\mathbf{K}_{S}+\mathbf{T}^{\left(
S\right)  }+\mathbf{T}^{\left(  D\right)  }\right)  +\frac{1}{4}%
\mathbf{D}\times \left( \mathbf{W}+\mathbf{K}_{D} \right) ,%
\end{equation}%
and

\begin{equation}
\frac{d \mathbf{D}\left(  t^{\prime}\right)  }{d t^{\prime}%
}=\mathbf{D}\times\left(  \mathbf{H}+\mathbf{K}_{S}+\mathbf{T}^{\left(
S\right)  }\right)  +\mathbf{S}\times\left(  \mathbf{K}_{D}+\mathbf{W}\right)
+\mathbf{D}\times\mathbf{T}^{\left(  D\right)  } \; .%
\end{equation}

\bigskip

\noindent We observe that the additive terms on the right 
now become K-dependent. \ Hence 
the fluctuations of the magnitude of the magnetization are 
anisotropy-dependent. \ The equation of motion for $\mathbf{D}$ shows that anisotropy
dependence can be in the precessional term and hence can be recovered 
even in the classical limit beyond a linear-response 
approach. \ We next show that this indeed the case.  

\bigskip 
\indent  From Eq.$\left(\ref{anisotropy} \right)$, the phase of the path integral, which
we denote by $\mathbf{\Gamma}$, is identified with the
effective action for the spin system. \ The procedure to find 
correlation functions is standard \cite{Schwinger}. To this 
action  we add two independent external
sources $\mathbf{Q}_{1}$ , $\mathbf{Q}_{2}$. These sources will then be
coupled to the fields\textbf{\ }$\mathbf{S}_{1}$ and $\mathbf{S}_{2}$,
respectively, and will be used to generate  two-point correlation functions
for the fields $\mathbf{S}$ and $\mathbf{D}$. \ Since in the
following we restrict ourselves to equilibrium properties, the sources 
$\mathbf{Q}_{1}$ and $\mathbf{Q}_{2}$ will be taken as equal, but 
still arbitrary. This is equivalent to
making  the external field $\mathbf{H}$ arbitrary. \ Hence 
$\mathbf{H}$ will be used instead to generate 
the correlation functions. \ Since we are 
solely interested in how 
the correlation functions depend on the anisotropy constant $K$, 
at the end we take the limit $\mathbf{H} \rightarrow 0 $. \ We 
also restrict the discussion to small deviations of the 
magnetization from the $z-$axis.

\bigskip

\indent First, we define a new functional $\mathbb{Z}$ of the 
external field $\mathbf{H}$,

\begin{eqnarray}
\mathbb{Z}\left[  \mathbf{H}\right]     &= & \int\mathfrak{D}\mathbf{S}_{f}%
\mathfrak{D}\mathbf{S}_{f^{\prime}}\int\mathfrak{D}\mathbf{\Omega}%
_{1}\mathfrak{D}\mathbf{\Omega}_{2}\langle\mathbf{\Omega}_{1}|\rho_{s}%
(0)|\Omega_{2}\rangle\\
&  &\times\int\mathfrak{D}\mathbf{S}_{1}\mathfrak{D}\mathbf{S}_{2}\; \exp\left[
\; \mathbf{\Gamma} \left[  \; \mathbf{H}\; \right] \; \right] . \nonumber
\end{eqnarray}%
This functional is defined in such a way that its variations with 
respect to $\mathbf{H}$ generate all correlation functions of the 
fields $\mathbf{S}$ and $\mathbf{D}$ at equilibrium. \ The average value 
of the field $\mathbf{D}$ is clearly given by
\begin{equation}
{\left.  \frac{1}{\mathbb{Z}}\frac{\delta\mathbb{Z}}{\delta H_{\alpha}\left(
t\right)  }\right|}_{\mathbf{H}\longrightarrow0}=i\left\langle D_{\alpha
}\left(  t\right)  \right\rangle .
\end{equation}%
The two-point irreducible correlation
function is similarly given by

\begin{eqnarray}
{\left. \frac{1}{\left(  i\right)  ^{2}}\frac{\delta^{2}{ln}\mathbb{Z}}{\delta H_{\alpha
}\left(  t\right)  \delta H_{\beta}\left(  t^{\prime}\right)  }  \right|}_{\mathbf{H}\longrightarrow0}& =\langle
D_{\alpha}\left(  t\right)  D_{\beta}\left(  t^{\prime}\right) \rangle- 
\langle
D_{\alpha}\left(  t\right) \rangle \langle D_{\beta}\left(  t^{\prime}\right)  \rangle \\
& \equiv \mathcal{G}_{\alpha\beta}^{-1}\left(  t-t^{\prime}\right) . \nonumber
\end{eqnarray}%
From the equations of motion of the fields $\mathbf{S}$ and $\mathbf{D}$,
we deduce the full equation of motion of the function $\mathcal{G}^{-1}_{\alpha\beta}$,

\begin{eqnarray}
\frac{\partial}{\partial t}\mathcal{G}^{-1}_{\alpha\gamma}\left(  t-t^{\prime}\right)   &
=& \epsilon^{\alpha\beta\lambda}\mathcal{G}_{\beta\gamma}\left(  t-t^{\prime}\right)
H_{\lambda}  + \epsilon^{\alpha\beta\lambda}\left(  2K\right)\mathcal{G}_{\beta\gamma}
\left(  t-t^{\prime}\right)
  \delta_{\lambda3}\langle S_{\lambda}\left(  t\right)
\rangle
\label{correlation2}\\
&& +   \epsilon^{\alpha\beta\lambda}\left(  2K\right)  \delta_{\lambda3}%
\mathcal{G}_{\lambda\gamma}\left(  t-t^{\prime}\right)  \langle S_{\beta}\left(
t^{\prime}\right)  \rangle   +i \epsilon^{\alpha\beta\lambda}\mathcal{G}_{\beta\gamma}\left(  t-t^{\prime
}\right)  \langle T_{\lambda}^{s}\left(  t\right)  \rangle\nonumber\\    
&& + i \epsilon^{\alpha\beta\lambda}\langle S_{\beta}\left(  t^{\prime
}\right)  \rangle\frac{\delta}{i\delta H_{\gamma}\left(  t\right)  }\langle
W_{\lambda}\left(  t\right)  \rangle \; ,\nonumber
\end{eqnarray}%
where $\epsilon^{\alpha\beta\gamma}$ is the antisymmetric unit tensor.
In the following the quantum terms are neglected and we look for corrections
to $\mathcal{G}^{-1}$ due only to the anisotropy term. \ We solve 
Eq.$\left( \ref{correlation2}\right)$ by iteration, 
starting from the free $\mathcal{G}^{-1}$ propagator. \ We find that the off-diagonal terms are
 now non-zero and depend explicitly on $K$. \ For the 1-2 element of 
$\mathcal{\mathcal{G}}^{-1}$,
we find that

\begin{equation}
\frac{\partial}{\partial t}\mathcal{G}_{12}^{-1}
\left(  t-t^{\prime}\right)  =A\frac{K}%
{T}\delta\left(  t-t^{\prime}\right)
\end{equation}%
where $A$ is a constant proportional to the 
magnetization. \ To obtain this equation we  also assumed
that the average of the vector $\mathbf{D}$ is zero. \ This assumption is 
due to the fact that the average of the classical stochastic field, 
$\mathbf{\xi}$, is also 
taken to be zero.

We define ${\widetilde{\mathcal{G}}}^{-1}$, the 
Fourier Transform of $\mathcal{G}^{-1}$, using
\begin{equation}
\mathcal{G}_{12}^{-1}\left( t-t^{\prime}\right)=\frac{1}{2\pi}\int 
d\omega \; 
e^{i\omega
\left(  t-t^{\prime}\right)  }
\widetilde{\mathcal{G}}_{12}^{-1}\left(  \omega\right) \; .
\end{equation}%
Solving in frequency Fourier space, we find that 
\begin{equation}
\widetilde{\mathcal{G}}_{12}^{-1}\left(  \omega\right) 
=\frac{-iAK/T}{\alpha\omega} \; .
\end{equation}
The other terms behave similarly as a function of 
$K/ T$. \ Since within the above assumptions, the correlation functions 
of the classical thermal field are the inverse of those of
 $\mathbf{D}$-field, we 
conclude  
that thermal fluctuations and anisotropy behave  oppositely.
The less the anisotropy the higher the level of 
fluctuations. \ Higher order 
corrections can similarly be calculated in this manner.

\bigskip

\section{ \bigskip CONCLUSION}

Using coherent states and a simple quantum 
mechanical model for a single large spin 
particle, we have shown that a generalized form of the Landau-Lifshitz
 equation can  
be recovered in the limit of high temperature. \ In this case the 
damping constant provides relaxation to the local equilibrium 
state. \ We  have 
also shown how fluctuations give two different contributions to the 
magnetization. \ One contribution is magnitude conserving and the 
other is not. \ We 
  derived generalized equations for the 
magnetization that include non-local effects in the dissipation
term and go beyond the simple linear-response approach. \ An important 
immediate result of this work is the dependence of fluctuations 
on the anisotropy of the system.  \ The LLGB equation 
is clearly inadequate in this 
respect. \ However these deficiencies can be corrected by using the right 
correlation functions for the fluctuations. \ Changing the damping
 to a tensor quantity in the LLG equation \cite{bertram2}
to account 
for noise correctly is then not needed. \ More complicated couplings, other 
than  the linear coupling considered here, naturally induce a tensor 
character for the relaxation. \ Most of these 
conclusions are true even in the classical limit. \ Garcia-Palacios 
\cite{Garcia} treated 
similar questions at the classical level and our results agree in that 
limit. \ Generalizing our model to the anisotropic case does not change 
these conclusions. \ In the linear approximation, the results of 
Safonov and Bertram \cite{bertram2} are reproduced from our 
results by taking the vector $\mathbf{D}$ to be the stochastic field. 
\ However the results of Smith and Arnett \cite{NeilSmith} are found by taking
the time derivative of the vector  $\mathbf{D}$ to be the random 
force. \ Clearly, as seen from Eq.$\left(\ref{llg2}\right) $, 
in this latter case the correlation functions of the 
random force can not be taken to be independent of the dynamics of the 
system. \ An isotropic fluctuation is simply not accurate
 enough for an asymmetric
Hamiltonian. \ More details on this comparison will be communicated elsewhere.

\bigskip

\section*{ACKNOWLEDGMENTS }
We would like to thank L. Benkhemis, R.W. Chantrell, 
W.N.G. Hitchon for critical
reading of the manuscript. Discussions with O. Chubykalo, C. Goebel,
A. Lyberatos, V. Safonov and 
X. Wang were also very enlightening.

\end{document}